\documentstyle[twoside,epsf,psfig]{article}

\catcode`\@=11
\long\def\@makefntext#1{
\protect\noindent \hbox to 3.2pt {\hskip-.9pt  
$^{{\eightrm\@thefnmark}}$\hfil}#1\hfill}		

\def\@makefnmark{\hbox to 0pt{$^{\@thefnmark}$\hss}}	
	
\def\ps@myheadings{\let\@mkboth\@gobbletwo
\def\@oddhead{\hbox{}
\rightmark\hfil\eightrm\thepage}   
\def\@oddfoot{}\def\@evenhead{\eightrm\thepage\hfil
\leftmark\hbox{}}\def\@evenfoot{}
\def\sectionmark##1{}\def\subsectionmark##1{}}

\oddsidemargin=\evensidemargin
\newcounter{sectionc}\newcounter{subsectionc}\newcounter{subsubsectionc}
\renewcommand{\section}[1] {\vspace{12pt}\addtocounter{sectionc}{1} 
\setcounter{subsectionc}{0}\setcounter{subsubsectionc}{0}\noindent 
	{\tenbf\thesectionc. #1}\par\vspace{5pt}}
\renewcommand{\subsection}[1] {\vspace{12pt}\addtocounter{subsectionc}{1} 
\setcounter{subsubsectionc}{0}\noindent 
{\bf\thesectionc.\thesubsectionc. {\kern1pt \bfit #1}}\par\vspace{5pt}}
\renewcommand{\subsubsection}[1] {\vspace{12pt}\addtocounter{subsubsectionc}{1}
	\noindent{\tenrm\thesectionc.\thesubsectionc.\thesubsubsectionc.
	{\kern1pt \tenit #1}}\par\vspace{5pt}}
\newcommand{\nonumsection}[1] {\vspace{12pt}\noindent{\tenbf #1}
	\par\vspace{5pt}}

\newcounter{appendixc}
\newcounter{subappendixc}[appendixc]
\newcounter{subsubappendixc}[subappendixc]
\renewcommand{\thesubappendixc}{\Alph{appendixc}.\arabic{subappendixc}}
\renewcommand{\thesubsubappendixc}
	{\Alph{appendixc}.\arabic{subappendixc}.\arabic{subsubappendixc}}

\renewcommand{\appendix}[1] {\vspace{12pt}
        \refstepcounter{appendixc}
        \setcounter{figure}{0}
        \setcounter{table}{0}
        \setcounter{lemma}{0}
        \setcounter{theorem}{0}
        \setcounter{corollary}{0}
        \setcounter{definition}{0}
        \setcounter{equation}{0}
        \renewcommand{\thefigure}{\Alph{appendixc}.\arabic{figure}}
        \renewcommand{\thetable}{\Alph{appendixc}.\arabic{table}}
        \renewcommand{\theappendixc}{\Alph{appendixc}}
        \renewcommand{\thelemma}{\Alph{appendixc}.\arabic{lemma}}
        \renewcommand{\thetheorem}{\Alph{appendixc}.\arabic{theorem}}
        \renewcommand{\thedefinition}{\Alph{appendixc}.\arabic{definition}}
        \renewcommand{\thecorollary}{\Alph{appendixc}.\arabic{corollary}}
        \renewcommand{\theequation}{\Alph{appendixc}.\arabic{equation}}
        \noindent{\tenbf Appendix \theappendixc #1}\par\vspace{5pt}}
\newcommand{\subappendix}[1] {\vspace{12pt}
        \refstepcounter{subappendixc}
        \noindent{\bf Appendix \thesubappendixc. {\kern1pt \bfit #1}}
	\par\vspace{5pt}}
\newcommand{\subsubappendix}[1] {\vspace{12pt}
        \refstepcounter{subsubappendixc}
        \noindent{\rm Appendix \thesubsubappendixc. {\kern1pt \tenit #1}}
	\par\vspace{5pt}}

\newcommand{\textlineskip}{\baselineskip=13pt}
\newcommand{\smalllineskip}{\baselineskip=10pt}

\newcommand{\copyrightheading}[1]
	{\vspace*{-2.5cm}\smalllineskip{\flushleft
	{\footnotesize International Journal of Modern Physics D, #1}\\
	{\footnotesize \copyright\kern2pt World Scientific Publishing
	 Company}\\
	 }}

\newcommand{\publisher}[2]{{\begin{center}\footnotesize\smalllineskip 
	Received #1\\
	Revised #2
	\end{center}
	}}

\def\abstracts#1#2#3{{
	\centering{\begin{minipage}{4.5in}\footnotesize\baselineskip=10pt
	\parindent=0pt #1\par 
	\parindent=15pt #2\par
	\parindent=15pt #3
	\end{minipage}}\par}} 

\renewenvironment{thebibliography}[1]
        {\frenchspacing
	 \ninerm\baselineskip=11pt
         \begin{list}{\arabic{enumi}.}
        {\usecounter{enumi}\setlength{\parsep}{0pt}     
	 \setlength{\leftmargin 12.7pt}{\rightmargin 0pt}
         \setlength{\itemsep}{0pt} \settowidth
	{\labelwidth}{#1.}\sloppy}}{\end{list}}

\newcounter{itemlistc}
\newcounter{romanlistc}
\newcounter{alphlistc}
\newcounter{arabiclistc}

\newcommand{\fcaption}[1]{
        \refstepcounter{figure}
        \setbox\@tempboxa = \hbox{\footnotesize Fig.~\thefigure. #1}
        \ifdim \wd\@tempboxa > 5in
           {\begin{center}
        \parbox{5in}{\footnotesize\smalllineskip Fig.~\thefigure. #1}
            \end{center}}
        \else
             {\begin{center}
             {\footnotesize Fig.~\thefigure. #1}
              \end{center}}
        \fi}

\newcommand{\tcaption}[1]{
        \refstepcounter{table}
        \setbox\@tempboxa = \hbox{\footnotesize Table~\thetable. #1}
        \ifdim \wd\@tempboxa > 5in
           {\begin{center}
        \parbox{5in}{\footnotesize\smalllineskip Table~\thetable. #1}
            \end{center}}
        \else
             {\begin{center}
             {\footnotesize Table~\thetable. #1}
              \end{center}}
        \fi}

\def\@citex[#1]#2{\if@filesw\immediate\write\@auxout
	{\string\citation{#2}}\fi
\def\@citea{}\@cite{\@for\@citeb:=#2\do
	{\@citea\def\@citea{,}\@ifundefined
	{b@\@citeb}{{\bf ?}\@warning
	{Citation `\@citeb' on page \thepage \space undefined}}
	{\csname b@\@citeb\endcsname}}}{#1}}

\newif\if@cghi
\def\cite{\@cghitrue\@ifnextchar [{\@tempswatrue
	\@citex}{\@tempswafalse\@citex[]}}
\def\citelow{\@cghifalse\@ifnextchar [{\@tempswatrue
	\@citex}{\@tempswafalse\@citex[]}}
\def\@cite#1#2{{$\null^{#1}$\if@tempswa\typeout
	{IJCGA warning: optional citation argument 
	ignored: `#2'} \fi}}

\def\pmb#1{\setbox0=\hbox{#1}
	\kern-.025em\copy0\kern-\wd0
	\kern.05em\copy0\kern-\wd0
	\kern-.025em\raise.0433em\box0}

\def\fnt#1#2{\footnotetext{\kern-.3em
	{$^{\mbox{\scriptsize #1}}$}{#2}}}

\def\fpage#1{\begingroup
\voffset=.3in
\thispagestyle{empty}\begin{table}[b]\centerline{\footnotesize #1}
	\end{table}\endgroup}

\def\runninghead#1#2{\pagestyle{myheadings}
\markboth{{\protect\footnotesize\it{\quad #1}}\hfill}
{\hfill{\protect\footnotesize\it{#2\quad}}}}
\font\tenrm=cmr10
\font\tenit=cmti10 
\font\tenbf=cmbx10
\font\bfit=cmbxti10 at 10pt
\font\ninerm=cmr9

\font\eightrm=cmr8

\def\qed{\hbox{${\vcenter{\vbox{	          
   \hrule height 0.4pt\hbox{\vrule width 0.4pt height 6pt
   \kern5pt\vrule width 0.4pt}\hrule height 0.4pt}}}$}}

\def\PRL{\em Phys. Rev. Lett.}
\def\PRD{{\em Phys. Rev.} D}

\def\AA{\em Astron. Astrophys.}
\begin{document}
\setlength{\textheight}{7.7truein}    

\runninghead{About the detection of gravitational wave bursts} {About the detection of gravitational wave bursts}

\normalsize\textlineskip
\thispagestyle{empty}
\setcounter{page}{1}

\copyrightheading{}		

\vspace*{0.88truein}

\fpage{1}
\centerline{\bf ABOUT THE DETECTION OF GRAVITATIONAL WAVE BURSTS}
\vspace*{0.37truein}
\centerline{\footnotesize T.PRADIER, N.ARNAUD, M.-A. BIZOUARD, F. CAVALIER, M. DAVIER, P.HELLO}
\vspace*{0.015truein}
\centerline{\footnotesize\it Laboratoire de l'Acc\'el\'erateur Lin\'eaire}
\centerline{\footnotesize\it B.P. 34 B\^atiment 208, Campus d'Orsay}
\centerline{\footnotesize\it 91898 Orsay Cedex, France}
\baselineskip=10pt
\vspace*{0.225truein}
\publisher{(received date)}{(revised date)}

\vspace*{0.21truein}
\abstracts{Several filtering methods for the detection of gravitational wave bursts in interferometric detectors are presented. These are simple and fast methods which can act as online triggers. All methods are compared to matched filtering with
the help of a {\it figure of merit} based on the detection of  supernovae signals simulated by 
Zwerger and M\"uller.}{}{}



\vspace*{1pt}\textlineskip	
\section{Introduction}	
\vspace*{-0.5pt}
\noindent
Supernovae have been historically the first envisaged sources of gravitational waves (GW). 
Although binary inspirals or even periodic
GW emitters like pulsars seem to be nowadays more promising sources, impulsive sources 
of GW such as supernovae should also be considered
in the data analysis design of interferometric detectors currently under construction (LIGO, VIRGO). Impulsive GW sources are typically collapses of massive stars, leading to the 
birth of a neutron star 
(type II supernova) \cite{bona,zwerger,rampp} or of a black hole \cite{stark}; 
mergers of compact binaries can also be considered as impulsive sources \cite{ruffert}. 

The problem with such sources is that the emitted waveforms are very poorly predicted, 
unlike the binary inspirals.
As a consequence, this forbids the use of matched filtering for the detection of GW 
bursts.
The filtering of such bursts should therefore be as general and robust as possible and with 
minimal {\it a priori} assumptions on the waveforms. A drawback
is of course that such filters will be sensitive to non-stationary noise as well as to GW bursts;
spurious events, e.g. generated by transient noise, should be eliminated afterwards when
working in coincidence with other detectors.
But, on the other hand, burst filters could help to identify and understand these noise sources, which would
be useful especially during the commissioning phase of the detector.

All the filters presented here are dedicated to GW bursts detection and are compared by studying their performance
to detect a reference sample of GW burst signals, numerically computed by Zwerger and M\"uller 
(ZM).\cite{zwerger}

Throughout the following, we assume that the detector noise 
is white, stationary and Gaussian with zero mean. For numerical 
estimates, we chose the flat (amplitude) spectral density to 
be $h_n \simeq 4\times10^{-23} /\sqrt{\mathrm Hz}$ and the sampling 
frequency $f_s \simeq 20$ kHz, so the standard deviation of the noise is
$\sigma_n = h_n \sqrt{f_s/2} \sim 4\times10^{-21}$. The value chosen for $h_n$ corresponds 
approximately to the minimum of the sensitivity curve of the 
VIRGO detector \cite{virgosens};
around this minimum, the sensitivity is rather flat, in the range [200 Hz,1kHz], which is precisely the range 
of interest for the gravitational wave bursts
we are interested in. This validates then our assumption of a white noise ; 
otherwise, we can always assume that the detector output
 has been first whitened by a suitable filter \cite{cuo}.

\section{General filters}
\vspace {-0.2cm}
\subsection{Filters based on the autocorrelation}
\noindent
The noise being whitened, its autocorrelation is ideally a Dirac function and in practice vanishes outside of zero. The autocorrelation of the data $x(t)$ 
\begin{equation}
A_x(\tau) = \int x(t) x(t+\tau) dt
\end{equation}
should then reveal
the presence of some signal (which is surely correlated). The information contained in the autocorrelation
function can be extracted in different ways. We have studied two of them and built so two non-linear 
filters. The first one computes the maximum of $A_x(\tau)$ and has already been described in \cite{nous}. In the following, we will refer to this filter as the Norm Filter (NF). A similar approach has been developed independently
by Flanagan and Hughes in the context of the detection of binary black hole mergers \cite{flana}.

Another possibility is to look at the norm of the autocorrelation function (NA Filter) :
\begin{equation}
||A|| = \sqrt{ {1 \over N} \sum_{k=2}^N A(k)^2},
\end{equation}
where $A(k)$ denotes the discrete autocorrelation of $N$ data $x_i$. The sum is here initiated at the second bin according to the fact
that the noise (uncorrelated) contributes essentially to the first bin. 
Note that the only parameter for these two filters is the window size $N$.
The behavior of the NA filter with noise only is not known analytically and its characteristics (mean and standard deviation) have to be found numerically (adding some complexity
to this filtering method).

\subsection{The Bin Counting method}
\noindent
This filter (BC) computes the number of bins in a window of size N whose value exceeds some threshold $s \times \sigma_n$. The threshold $s$ is chosen by maximizing the signal to noise ratio (SNR) when detecting the signals of the ZM catalogue (for more details, see \cite{nous}).

\subsection{Linear Fit Filters}
\noindent
This filter fits the data to a straight line in a window of size $N$. If the data are pure white noise with zero mean the slope and the offset of the fitted line are zero on average, so this can well discriminate between the two cases : only noise or noise+signal.
The slope and the offset of the fit can easily be computed as a function of time and of the $x_i$ (see \cite{slope}).

In fact, the slope and the offset are two {\it correlated} random variables. By computing their Covariance Matrix, one reduces them to two {\it uncorrelated }normal random variables, $X_+$ and $X_-$, with

\begin{equation}
X_{\pm} = \left( X \pm Y \over \sqrt{2(1 \pm \alpha)} \right),  \mbox{with  } \alpha = -\sqrt{{ 3\over2} \left({N + 1\over 2N + 1 }\right)} = COV(X,Y)  
\end{equation}

where X = SNR(slope) = $|slope|/\sigma_{slope}$, Y = SNR(offset).
Finally, the sum $X_+^2 + X_-^2$ is a $\chi^2$ like random variable and gives us a new filter, ALF (Advanced Linearfit Filter).

\subsection{The Peak Correlator}
\noindent
Filtering by correlating the data with peak (or pulse) templates is justified by the fact that simulated supernovae GW signals exhibit one
(or more) peaks. The pulse templates have been built from truncated Gaussian functions. The method and results are explained in \cite{nous}.

\section{Performance and efficiency of the filters}
\vspace {-0.2cm}
\subsection{Definition of a false alarm rate}
\noindent
We {\it arbitrarily} set the  false alarm rate for each of the filters to be $10^{-6}$ (72 false alarms per hour for a sampling
frequency $f_s=20$ kHz). This high rate is required because the signals we look for are very weak. False alarms will be discarded later when working in coincidence.

\subsection{The Zwerger and M\"uller Catalogue}
\noindent
The catalogue of Zwerger and M\"uller \cite{webSN} contains 78 
gravitational-wave signals.  Each of them corresponds to a particular 
set of parameters (e.g initial distribution of angular momentum). All the signals are computed for a source located at 10\,Mpc.
We can then re-scale the waveforms in order to locate the source at any distance $d$.

Since the signal waveforms are here known, we can explicitly derive the 
optimal SNR provided by the Wiener filter matched to each of them, and 
then compute the
maximal distance of detection. We will then be able to build a benchmark 
for the different filters by comparing their results (detection distances) to the results
of the Wiener filter (we consider here optimally polarized GW's, along
the interferometer arms).

The mean distance obtained for the Wiener Filter, averaged over all the signals, is about $\bar{d}_{\mathrm opt} \simeq 25.4$\,kpc, 
which is of the order of the diameter of the Milky Way. 

\subsection{Estimating a filter power}
\noindent
The optimal (Wiener) filtering
allows to detect the $i^{\mathrm th}$ signal in the Catalogue emitted by a source located up to a distance $d_i^{\,({\cal W})}$. Similarly,
a filter ${\cal F}$ is able to detect the same signal up to a distance $d_i^{\,({\cal F})}$; of course $d_i^{\,({\cal F})}$ is averaged
over many noise realizations (about 1000) in a Monte Carlo simulation. The detection performance of the filter ${\cal F}$ for this signal is simply defined as the distance of detection relative to
the optimal distance of detection : $d_i^{\,({\cal F})}/d_i^{\,({\cal W})}$. The global performance of ${\cal F}$ is then estimated as the detection performance averaged over all the waveforms of the catalogue:

\noindent
\begin{equation}
\rho= {1 \over 78} \sum_{i=1}^{78} {d_i^{\,({\cal F})} \over d_i^{\,({\cal W})}}.
\end{equation}

For a given filter, and a given source located at a distance $d$, one can also evaluate a {\it detection efficiency} $\epsilon$, which is the number of detections $n$ over the total number of noise realisations ${\cal N}$. This efficiency (averaged over all the signals of the catalogue)  will characterize the {\it practical} behaviour of the filter.

\subsection{Comparison of the filtering methods : Performance}
\noindent
The results for the different filters are reported in Table \ref{tab}.
We also give the average distance of 
detection $\bar{d}= {1 \over 78} \sum_{i=1}^{78} d_i^{({\cal F})}$ for all the filters.

\vspace {-0.5cm}
\begin{table}[ht]
\caption{Performance of the different filters. L means linear filter and NL means
non-linear filter.\label{tab}}
\vspace{0.4cm}
\begin{center}
\begin{tabular}{|c|c|c|c|c|c|c|c|c|}
\hline
Filter& Optimal & NF& NA & BC & PC & $X_+$ &$X_-$ & ALF \\ 
\hline
$\bar{d}$ (kpc)& 25.4 & 11.5 & 11.4 & 10.9 & 18.5 & 21.6 & 22.2 & 22.4\\
$\rho$                 & 1      &0.46&0.47&0.43& 0.73 & 0.79 & 0.80 & 0.81\\
Linearity &L&NL&NL&NL&L&L&L&NL \\ 
\hline
\end{tabular}
\end{center}
\end{table}

The three first filters NF, NA and BC (all non-linear) have a performance slightly below one half, while the PC
have a performance greater than 0.7. Both $X_+$ (or $X_-$) and ALF can reach a performance around 0.8. Note that ALF has been in fact implemented with a sampling of 20 different window sizes,
sufficient to cover the variety of signals (with a non-significant loss of generality). If implemented with a single window size, as the other filters NF, NA and BC,
its performance decreases down to 0.72.

\subsection{Comparison of the filtering methods : Detection efficiency}
\noindent
For the Wiener filter, one can show that the mean efficiency is roughly $50 \%$ for signals located at a distance $d_i^{\,({\cal W})}$. If $\bar{\rho}$ is the mean performance of a given filter, one would expect a $50 \%$ detection efficiency at $d_i^{\,({\cal F})} \simeq \bar{\rho} \times d_i^{\,({\cal W})}$. In fact, such efficiency is reached for a smaller distance.

If one defines the {\it effective performance } $\rho_{eff}$ as the ratio $d_i^{\,({\cal F})}/ d_i^{\,({\cal W})}$ for which the detection efficiency is about $50 \%$, then $\rho_{eff} \simeq 0.74$ for ALF and $X_-$ and $\rho_{eff} \simeq 0.71$ for $X_+$. This definition gives an idea of the efficiency one can reach in practice, and has to be taken into account when choosing between different online triggers.

%
%
%
\section{Conclusion}
\noindent
We have discussed several filters to be used as triggers for detecting GW bursts in interferometric detectors.
They are all sub-optimal but their performance is close to the one obtained with the Wiener Filter.

Concerning the detection of Zwerger-M\"uller-like signals, we note that
none of the BC, NF and NA filters is efficient enough to cover  the whole Galaxy in average (but their window sizes have not yet been optimized), contrary to ALF and PC 
(and optimal) filters.
A  few signals can be detected at distances beyond 50\,kpc, 
the distance of the Large Magellanic Cloud (LMC). 
It is clear that this class of signals will be detected by the first generation interferometric 
detectors such as VIRGO only if the supernovae occur inside our Galaxy or in the very
close neighbourhood.

Finally, all the filters studied here can be implemented on line without problem, due to use of FFT's (for the NA and the PC)
or to simple recursive relations between filter outputs in successive windows (NF,BC or ALF).

Correlations and coincidences between those filters are under study in order to either reduce background (hence a quantifiable loss of signal) or lower detection thresholds (hence a gain of a few $\%$ in performance and efficiency).
%

\nonumsection{References}

\end{document}